\documentclass[10pt, final, journal, letterpaper, twocolumn]{IEEEtran}

\usepackage{amsfonts}
\usepackage[dvips]{graphicx}
\usepackage{times}
\usepackage{cite}
\usepackage{amsmath}
\usepackage{cases}
\usepackage{array}
\usepackage{dsfont}
\usepackage{amssymb}

\usepackage{stfloats}
\usepackage{slashbox}
\usepackage{graphicx}
\usepackage{footnote}
\usepackage{booktabs}
\usepackage{array}
\usepackage{bm}
\usepackage{color}

\begin{document}

\title{Coordinated Multicast Beamforming in Multicell Networks}

\author{\IEEEauthorblockN{Zhengzheng Xiang, Meixia Tao, \IEEEmembership{Senior Member, IEEE}, and Xiaodong Wang, \IEEEmembership{Fellow, IEEE}}
\thanks{Manuscript received December 29, 2011; revised April 22, 2012, and July 26, 2012; accepted September 19, 2012. The associate editor coordinating the
review of this paper and approving it for publication was Prof. Sezgin Aydin.}
\thanks{Z. Xiang and M. Tao are with the Dept. of Electronic Engineering, Shanghai Jiao Tong University, P. R. China. Email:\{7222838, mxtao\}@sjtu.edu.cn.}
\thanks{X. Wang is with the Department of Electrical Engineering at Columbia University, New York, USA. Email:
wangx@ee.columbia.edu.}
\thanks{This work is supported by the Joint Research Fund for Overseas Chinese,
Hong Kong and Macao Young Scholars under grant 61028001, the National
973 project under grant 2012CB316100, and  the New Century Excellent
Talents in University (NCET) under grant NCET-11-0331".
Please also mention that part of this work will be presented in GLOBECOM
2012.}
}
 \maketitle

\begin{abstract}
We study physical layer multicasting in multicell
networks where each base station, equipped with multiple antennas,
transmits a common message using a single beamformer to multiple
users in the same cell. We
investigate two coordinated beamforming designs: the
quality-of-service (QoS) beamforming and the max-min SINR (signal-to-interference-plus-noise ratio) beamforming.
The goal of the QoS beamforming is to minimize the total power
consumption while guaranteeing that received SINR at each user is above
a predetermined threshold. We present a necessary condition
for the optimization problem to be feasible. Then, based on the
decomposition theory, we propose a novel decentralized algorithm to
implement the coordinated beamforming with limited information
sharing among different base stations. The algorithm is guaranteed
to converge and in most cases it converges to the optimal solution.
The max-min SINR (MMS) beamforming is to maximize the minimum received SINR among all users under per-base station power constraints. We show that the MMS problem and a weighted peak-power minimization
(WPPM) problem are inverse problems. Based on this inversion relationship, we then propose an efficient
algorithm to solve the MMS problem in an approximate manner. Simulation results
demonstrate significant advantages of the proposed multicast
beamforming algorithms over conventional multicasting schemes.

\end{abstract}

\begin{IEEEkeywords}
Physical layer multicasting, coordinated beamforming, quality of service (QoS), max-min SINR (MMS), semidefinite programming (SDP).
\end{IEEEkeywords}

\section{Introduction}
With the rapid development of wireless communication technology,
various kinds of traditional data service, such as media streaming,
cell broadcasting and mobile TV, have been implemented in wireless
networks nowadays. As a result, wireless multicasting becomes a
central feature of the next generation cellular networks. Physical
layer multicasting with beamforming is a promising solution enabled by exploiting
channel state information (CSI) at the transmitter over the
traditional isotropic broadcasting. The problem of multicast
beamforming for quality-of-service (QoS) guarantee and for max-min
fairness was firstly considered in \cite{Sidiropoulos}. The similar
problem of multicasting to multiple cochannel groups was then
investigated in \cite{Karipidis}. The core problem of multicast
beamforming is essentially NP-hard \cite{Sidiropoulos}. Some other
issues such as outage analysis and capacity limits were studied in
\cite{Ntranos}, \cite{Park}. So far multicast beamforming  has been
included in the UMTS-LTE / EMBMS draft for next-generation cellular
wireless services \cite{Motorola}, \cite{Lozano}.

 In conventional wireless systems, signal processing is
performed on a per-cell basis. The intercell interference is treated
as background noise and minimized by applying a predesigned
frequency reuse pattern such that the adjacent cells use different
frequency bands. Due to the fast growing demand for high-rate
wireless multimedia applications, many beyond-3G wireless
technologies such as 3GPP-LTE and WiMAX have relaxed the constraint
on the frequency reuse such that the total frequency band is
available for reuse by all cells in the same cluster. However, this
will cause the whole system limited by the intercell interference.
Consequently, cooperative signal processing across the different
base stations  has been identified as a key technique to mitigate
intercell interference in the next-generation wireless systems.

The goal of this paper is to investigate multicast beamforming for
cooperative multicell networks. The base stations cooperate with
each other and design their transmit beamformers in a coordinated
manner. In general, there are two cooperation scenarios in multicell
networks. In the first scenario, different base stations are fully
cooperative and act as ``networked MIMO (multiple-input
multiple-output)" (e.g., \cite{Shamai},\cite{Dai}). Namely, they
coordinate in the signal level, i.e., data information intended for
different users in different cells is shared among the base
stations. Clearly, networked MIMO needs tremendous amount of
information exchange overhead. In contrast, in the second scenario,
the base stations are only required to coordinate at the beamforming
level which needs rather small information sharing (e.g., \cite{Jorswieck}).
In this paper, we focus on the latter case by only allowing beamforming level coordination.

We first formulate the problem of multicell multicast beamforming as
sum power minimization subject to the constraint that the received
signal-to-interference-plus-noise ratio (SINR) of each user is above
a threshold. This problem is referred to as QoS problem. It is known that the QoS problem  in
single-cell scenario is always feasible. However, this is not so in
multicell networks due to the intercell interference, especially
when the channel condition is bad or the SINR target is very
stringent. We first present a necessary condition  on the
optimization problem to be feasible. Then we propose a novel
distributed algorithm to solve the multicell multicast
cooperative beamforming design. More specifically, based on the
decomposition theory \cite{Palomar}, we introduce a set of interference constraint
parameters and decompose the original problem into several parallel
sub-problems. Since the sub-problems are nonconvex and NP-hard, each base station then
applies the semidefinite relaxation and independently solves its own sub-problem.
The interference parameters are updated based on a master problem. Simulation results show that the algorithm
converges in only several iterations.

In view of the user fairness issue, we also consider the multicell
multiast beamforming design with the goal of maximizing the minimum
SINR (MMS) among all users under individual power constraints. This
is called MMS problem. By linking the MMS problem with a weighted
peak power minimization (WPPM) problem, we propose an efficient
algorithm to find the near-optimal solution. Previously, the authors
in \cite{Jordan} \cite{Dartmann} also studied the multicell
multicast beamforming problem with the objective to maximize the
minimum SINR but subject to a total sum power constraint across all
base stations. Hence, the problem considered in \cite{Jordan}
\cite{Dartmann} is very similar to the multicast beamforming problem
in single-cell multi-group systems as in \cite{Karipidis}. Authors
in \cite{Dartmann1} also considered the similar problem under
per-base-station power constraint but allowed data sharing among
base stations, which is similar to the ``network MIMO" and requires a large amount of
information exchange overhead. Thus, our beamforming design for max-min fairness is more practical
as we consider individual peak power constraint on each base station and only allow coordination in
the beamforming level.

 The rest of the paper is organized as follows. In Section II, the system model for
multicell multicasting is presented. Section III considers the
beamformer design in the QoS problem and develops a novel
decentralized coordinated beamforming algorithm. In Section IV, we
consider the max-min SINR beamforming problem under individual base
station power constraints. Section V provides simulation results.
Concluding remarks are made in Section VI.

\textsl{Notations:} $\mathds{R}$ and $\mathds{C}$ denote the real and
complex spaces. The identity matrix is denoted as
$\textsl{\textbf{I}}$ and the all-one vector is denoted as ${\bf
1}$. For a square matrix $\textsl{\textbf{S}}$, $\textsl{\textbf{S}}
\succeq 0$ means that $\textsl{\textbf{S}}$ is positive
semi-definite. $[{\bf H}]_{i,j}$ denotes its element in the $i$th
row and $j$th column. $(\cdot)^\textsl{T}$, $(\cdot)^\textsl{H}$,
$(\cdot)^\dag$ and $\mbox{Tr}\{\cdot\}$ stand for transpose,
Hermitian transpose, Moore Penrose pseudoinverse and the trace
respectively. $|x|$ denotes the absolute value of the scalar $x$ and
$\|\bf{x}\|$ denotes the Euclidean norm of the vector $\bf{x}$.

\begin{figure}
\begin{centering}
\includegraphics[scale=0.45]{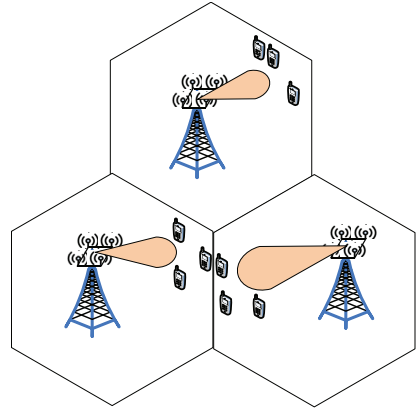}
\vspace{-0.1cm} \caption{Multicell multicast network.}
\label{fig:multicell_multicast_system}
\end{centering}
\vspace{-0.3cm}
\end{figure}

\section{System Model}
Consider a multicell multicast network comprising $N$ cells and $K$
mobile users per cell as shown in Fig. 1. The base station in each
cell is equipped with $N_t$ antennas and every mobile user has a
single antenna. Let ${\bf{h}}_{i,j,k}^H$ denote the frequency-flat
quasi-static $1\times N_t$ complex channel vector from the base
station in the $i$th cell to the $k$th user in the $j$th cell, and
$\textbf{w}_i$ denote the $N_t \times 1$ multicast
beamforming vector applied to the base station in the $i$th cell. We
define the complex scalar $s_i$ as the multicast information symbol
for the users in the $i$th cell. The discrete-time baseband signal
received by the $k$th user in the $i$th cell is given by
\begin{equation} \label{1}
    {y_{i,k}}={{\bf h}_{i,i,k}^{\emph{H}}}{{\bf
    w}_i}{s_i}+\sum^{N}_{j\neq i}{{\bf h}_{j,i,k}^{\emph{H}}}{{\bf
    w}_j}{s_j}+z_{i,k},
\end{equation}
where $z_{i,k}$ is the additive white circularly symmetric Gaussian
complex noise with variance $\sigma_{i,k}^2 /2$ on each of its real
and imaginary components. In (1), the second term is the intercell
interference.

Based on the received signal model in (1), the performance of each
user can be characterized by the output SINR, defined as
\begin{equation} \label{2}
    {\mbox{SINR}_{i,k}}=\frac{{|{\bf h}_{i,i,k}^{\emph H}{\bf w}_{i}|}^2}{\sum^{N}_{j\neq i}{|{\bf h}_{j,i,k}^{\emph H}{\bf
    w}_{j}|}^2+\sigma_{i,k}^2}.
\end{equation}
Notice that each user in the considered multicell multicast network
only suffers from the intercell interference, which is different
from multicell unicast systems where both inter-cell interference
and intra-cell interference exist.

In practical scenarios, the channels from a base station to different
users, which may or may not belong to a same cell, can be
correlated, especially when these users are in close proximity to each other.
For the users belonging to the same cell, we call the correlation between ${\bf h}_{i,j,k}^H$
and ${\bf h}_{i,j,k'}^H$ for $k\neq k'$ as the intracell-user
channel correlation. For the users from different cells, we call the correlation between
${\bf h}_{i,j,k}^H$ and ${\bf h}_{i,j',l}^H$ for
$j \neq j'$ as the intercell-user channel correlation.

\section{QoS Beamforming }
The coordinated QoS beamforming design is to minimize the total energy consumption of the system while maintaining a target SINR for all
users by properly designing the beamformers at each base station.
This is formulated as:

\begin{eqnarray}
\label{P_gamma_ob}{\bf P}({\bm \gamma}): &\min\limits_{\{{\bf
w}_i\}_{i=1}^{N}}& \sum\limits_{i=1}^{N}\|{\bf w}_i\|^2
\\ \label{P_gamma_cons} &\mbox{s.t.}&\frac{|{\bf h}_{i,i,k}^{\emph
H}{\bf w}_i|^2}{\sum\limits_{j\neq i}^{N}|{\bf h}_{j,i,k}^{\emph
H}{\bf w}_j|^2+\sigma^2_{i,k}}\geq \gamma_i, \forall i, k
\end{eqnarray}
where ${\bm \gamma}=[\gamma_1, \gamma_2,...,\gamma_N]^T$ is the
target SINR vector with each element $\gamma_i$ being the target
SINR value to be achieved by the users in the $i$th cell. Since the
base station transmits a common information in a multicast manner,
the information rate for the users within one cell is the same, and
hence we set a common SINR target for all the users in the same
cell.

\subsection{Feasibility analysis}
Due to the SINR constraints, the QoS problem in \eqref{P_gamma_ob}
is not always feasible, which is similar to the multiuser unicast scenario \cite{Wiesel}. In order to verify its feasibility, we need to show whether there exists a set of beamformers $\{{\bf w}_i
\}_{i=1}^{N}$ for a given ${\bm \gamma}$ such that
\begin{equation}
\min_{k}~\mbox{SINR}_{i,k}\geq \gamma_i, \forall i\in\{1,2,...,N\}
\end{equation}
For simplicity, the SINR targets for different cells are assumed to
be the same. Then the above condition boils down to the following
\begin{equation}
\min_{i,k}~\mbox{SINR}_{i,k}\geq \gamma
\end{equation}
For the $k$th user in each cell, we combine their channel vectors as
follows
\begin{equation}\label{H_k}
{\bf H}_k=\begin{bmatrix} {\bf h}_{1,1,k}^H & {\bf h}_{2,1,k}^H & \cdots & {\bf h}_{N,1,k}^H \\ {\bf h}_{1,2,k}^H & {\bf h}_{2,2,k}^H & \cdots & {\bf h}_{N,2,k}^H
\\\vdots & \cdots &~&~\\{\bf h}_{1,N,k}^H & {\bf h}_{2,N,k}^H & \cdots & {\bf h}_{N,N,k}^H \end{bmatrix},~k=1,2,...,K
\end{equation}
where each ${\bf H}_k$ is an $N\times(N\times N_t)$ matrix. The
following lemma provides a necessary condition for the QoS problem
to be feasible.

\textbf{Lemma 1}: Given the SINR target vector ${\bm
\gamma=\gamma\cdot{\bf 1}}$, if the problem \eqref{P_gamma_ob} is
feasible, then the SINR target $\gamma$ should satisfy the following
condition
\begin{equation}
\gamma \leq \min \left\{\frac{\mbox{rank}({\bf H}_1)}{{\emph
N}-\mbox{rank}({\bf H}_1)},\cdots,\frac{\mbox{rank}({\bf
H}_K)}{{\emph N}-\mbox{rank}({\bf H}_K)}\right\}
\end{equation}

\begin{proof}
Please refer to Appendix A.
\end{proof}

From Lemma 1, we can see that the intercell-user channel correlation has
a negative impact on its feasibility. More specifically, if some of the
rows in ${\bf H}_k$, say the $i$th row and
$i'$th row for $i\ne i'$, are correlated, which means that the channels between the $k$th user in the $i$th
cell and the $k$th user in the $i'$th cell are correlated, then the rank of ${\bf H}_k$ could be less than $N$ and as a result, the SINR threshold $\gamma$ will have a finite upper bound. On the other hand,
if the channels of users in different cells are independent, we
have that the matrix ${\bf H}_{k}$ is full rank with probability
one, then the corresponding upper bound is infinite, which means no
constraint on $\gamma$. Since the problem \eqref{P_gamma_ob} is
NP-hard \cite{Karipidis}, determining its feasibility is not an easy
job. The lemma gives a necessary condition from the perspective of
channel correlation. In the following, we only consider the problem
\eqref{P_gamma_ob} when it is feasible.

\subsection{Decentralized coordinated beamforming}
A desired feature of coordinated beamforming in multicell networks
is that the base station at each cell can implement its beamforming
design locally \cite{Dahrouj}, \cite{Zhang}. This is due to the constraint in practical systems that the backhaul
channel has limited capacity. In this subsection, we propose a
decentralized scheme for implementing the multicell multicast QoS
beamforming. It is assumed that each base station in the network
only has the channel knowledge of the mobile users within
its own cell.

The distributed algorithm to problem ${\bf P}(\bm \gamma)$ can not
be easily obtained primarily because all the beamformers are coupled
together in the constraint \eqref{P_gamma_cons}. According to the
decomposition theory \cite{Palomar} and inspired by \cite{Zhang}, we
introduce a set of slack variables $\Gamma_{i,j,k}$ denoting the
constraint of the interference from the $i$th base station to the
$k$th user in the $j$th cell. Then the problem $\bf P(\bm\gamma)$
can be reformulated as
\begin{eqnarray}
\label{P_gamma_Gamma_ob}{\bf P}(\bm \gamma, {\bf \Gamma}):&
\min\limits_{\{{\bf w}_i\}_{i=1}^{N}}&\sum_{i=1}^{N}{\|{\bf
w}_i\|}^2 \\ \label{P_gamma_Gamma_con1}& \mbox{s.t.} & \frac{{|{\bf
h}_{i,i,k}^{\emph H}{\bf w}_{i}|}^2}{\sum\limits_{j\neq
i}^{N}\Gamma_{j,i,k}+\sigma^2_{i,k}}\geq\gamma_i, \forall i, k
\\\label{P_gamma_Gamma_con2}& ~ & {|{\bf h}_{i,j,k}^{\emph H}{\bf
w}_i|}^2\leq \Gamma_{i,j,k}, \forall i,j\neq i,k
\end{eqnarray}
Here, the real-valued vector ${\bf \Gamma}$ is defined as follows
\begin{equation}{\bf
\Gamma}=(\Gamma_{1,2,1},...,\Gamma_{1,2,K},...
,\Gamma_{N,N-1,K})^{T}\in \mathds{R}^{\left(N\left(N-1\right)
K\right)\times 1}
\end{equation}

The introduction of ${\bf \Gamma}$ is similar to the concept of
interference temperature (IT) in cognitive radios and hence we refer
to it as IT vector. It is now observed that the constraints in
\eqref{P_gamma_Gamma_con1} are all decoupled. Unlike \cite{Zhang}, our problem is still
nonconvex and NP-hard. We cannot solve it by exploiting its dual problem since the strong duality does not hold.
In the following we take the semidefinite relaxation (SDR) approach, the optimality of which shall be discussed later in this section. Introducing new variables $\{ {\bf W}_i={\bf w}_i{\bf w}^{\emph H}_i \}_{i=1}^{N}$, the relaxed
problem of ${\bf P}({\bm\gamma}, {\bm\Gamma})$ becomes:
\begin{eqnarray}
\label{P_1_ob}{\bf P}_{1}(\bm\gamma, {\bf \Gamma}):
\min\limits_{\{{\bf W}_i\}_{i=1}^{N}}  \sum\limits_{i=1}^{N}
\mbox{Tr}\{{\bf W}_i\}~~~~~~~~~~~~~~~~~~~~~~~~~~~~~~\\
\label{P_1_con1}~ \mbox{s.t.}~~ {\mbox{Tr}\{{\bf H}_{i,i,k}{\bf
W}_i\}}\geq \gamma_i{{\bf e}_{i,i,k}^T{\bf
\Gamma}+\gamma_i\sigma^2_{i,k}} , \forall i, k
\\ \label{P_1_con2}  \mbox{Tr}\{{\bf H}_{i,j,k}{\bf W}_i\}\leq {\bf
e}_{i,j,k}^T{\bf \Gamma}, \forall i,j\neq i, k~~~~~
\\{\bf W}_i\succeq {\bf 0}~~~~~~~~~~~~~~~~~~~~~~~~~~~~~~~~~~~~~~
\end{eqnarray}
Here, for notation convenience we have introduced the
$\left(N\left(N-1\right)K\right)\times 1$ direction vectors ${\bf
e}_{i,j,k}$ and defined ${\bf H}_{i,j,k}\triangleq {\bf h}_{i,j,k}{\bf h}_{i,j,k}^H$.

For a pre-fixed IT vector $\bf \Gamma$, since the
constraints have been decoupled, problem ${\bf P}_1(\bm\gamma,{\bf
\Gamma})$ can be decomposed into $N$ parallel subproblems. The $i$th
subproblem is as follows
\begin{eqnarray}\label{P_sub}
\label{P_sub}{\bf P}^{\mbox{sub}}_{i}(\bm\gamma,{\bf \Gamma}):
~\min\limits_{{\bf W}_i} \mbox{Tr}\{{\bf
W}_i\}~~~~~~~~~~~~~~~~~~~~~~~~~~~~~~~~~~~
\\ \mbox{s.t.} ~~{\mbox{Tr}\{{\bf H}_{i,i,k}{\bf W}_i\}}\geq \gamma_i{{\bf
e}_{i,i,k}^T{\bf \Gamma}+\gamma_i\sigma^2_{i,k}} , \forall k
\\\mbox{Tr}({\bf H}_{i,j,k}{\bf W}_i)\leq {\bf e}_{i,j,k}^T{\bf \Gamma}, \forall j\neq i, k ~~~~~~\\{\bf
W}_i\succeq {\bf 0}~~~~~~~~~~~~~~~~~~~~~~~~~~~~~~~~~~~~
\end{eqnarray}
It is easily observed that in each subproblem, the base station only
needs the local channel state information. Specifically, the $i$th
cell's base station needs just the channel vectors ${\bf
h}_{i,j,k}^H,\forall j=1\cdots N,\forall k=1\cdots K$. Since it is
also convex, the optimal solution can be obtained efficiently.
Having solved the subproblems, we then define the master problem in
charge of updating the IT vector ${\bf \Gamma}$
\begin{equation}
{\bf P}^{\mbox{mas}}(\bm\gamma): ~\min_{\bf
\Gamma}~~P\left(\bm\Gamma\right)
\end{equation}
where $P\left(\bm\Gamma\right)=\sum\limits_{i=1}^NP_i^\star({\bf \Gamma})$ with $P_i^\star({\bf \Gamma})$ being the optimal solution of problem ${\bf P}^{\mbox{sub}}_{i}(\bm\gamma,{\bf \Gamma})$ for a given ${\bf
\Gamma}$. This master problem can be solved iteratively using a
subgradient projection method. Denote ${\bf g}\in \mathds{R}^{\left(N\left(N-1\right)
K\right)\times 1}$ as the global subgradient of $P\left(\bm\Gamma\right)$ at
${\bm \Gamma}$. The following theorem suggests that $\bf g$ can be obtained from each base station.

$\textbf{Theorem 1}:$ The global subgradient $\bf g$ of $P\left(\bm\Gamma\right)$
in the master problem ${\bf P}^{\mbox{mas}}(\bm\gamma)$ is given by
\begin{equation}
{\bf g}=\sum_{i=1}^{N}{\bf g}_i
\end{equation}
where ${\bf g}_i$ is the subgradient of $P_i^\star({\bf \Gamma})$.
\begin{proof}
Please refer to Appendix B.
\end{proof}

According to Theorem 1, every base station first solves its own
subproblem, gets the subgradient vector ${\bf g}_i$ and then broadcasts it
to other cells. Upon receiving all the $N$ subgradient vectors ${\bf g}_i$'s, the base
station in each cell will sum them up to get the subgradient ${\bf g}$ and update the IT
vector $\bf \Gamma$ as below
\begin{equation}\label{Gamma_update}
{\bf \Gamma}(n+1)=\left[{\bf \Gamma}(n)-\mu(n)\cdot\frac{{\bf
g}(n)}{\|{\bf g}(n)\|}\right]^+,
\end{equation}
where $n$ denotes the iteration index and $\mu$ is the step size.
Here, $[\cdot]^+$ denotes the projection onto the nonnegative
orthant. Since the problem ${\bf P}_{1}(\bm\gamma,{\bf \Gamma})$ is
convex, this distributed
algorithm is guaranteed to converge  and converge exactly to the
optimal solution of ${\bf P}_{1}(\bm\gamma,{\bf \Gamma})$. Just like
the steepest descent method, the choice of step size affects the
convergence properties of the iterative algorithm, such as the speed
and the accuracy. Here we just take the simple diminishing step,
i.e., $\mu(n)=s/{\sqrt n}$, where $s>0$ is the initial step size.

When the algorithm converges, each base station can get its own
beamformer from its resulting matrix ${\bf W}_i^\star$. We now discuss
how to extract the beamformer vector ${\bf w}_i$ from each ${\bf W}_i^\star$ and its optimality.
Since the original problem ${\bf P}(\bm \gamma)$ is NP-hard, generally there is no guarantee
that an algorithm for solving the relaxed SDP problem will give the desired rank-one solution.
If the ${\bf W}_i^\star$ is rank-one, then the base station applies the eigen-value decompsotion (EVD) to
${\bf W}_i^\star$ as ${\bf W}_i^\star=\lambda_i^\star{\bf
w}_i^\star{{\bf w}_i^\star}^H$ and takes ${\bf
w}_i=\sqrt{\lambda_i^\star}{\bf w}_i^\star$ as the optimal
beamformer. Otherwise, it first generates a set of candidate beamforming vectors
$\{{\bf w}_i^l\}$ by randomization method. Specifically, one can perform EVD on ${\bf
W}_i^\star$ to get ${\bf W}_i^\star={\bf U}_i{\bm
\Sigma}_i{\bf U}_i^H$ and then generate the $l$th candidate vector ${\bf w}_i^l$ as
${\bf w}_i^l={\bf U}_i{\bm \Sigma}_i^{1/2}{\bf v}_l$, where ${\bf
v}_l\sim \mathcal{CN}({\bf 0},\bf I)$. Then based on its own
subproblem ${\bf P}^{\mbox{sub}}_{i}(\bm\gamma,{\bf \Gamma})$, the
base station needs to do scaling to get the beamforming vector. The
scaling problem is formulated as below
\begin{eqnarray}
{\bf L}_{i}(\bm\gamma,{\bf \Gamma}): ~\min\limits_{\alpha_i}
~~\alpha_i~~~~~~~~~~~~~~~~~~~~~~~~~~~~~~~~~~~~~~~~~~~
\\ \mbox{s.t.} ~~{\alpha_i|{\bf h}_{i,i,k}^H{\bf w}'_i|^2}\geq\gamma_i{{\bf
e}_{i,i,k}^T{\bf \Gamma}+\gamma_i\sigma^2_{i,k}} , \forall k~~~~~~
\\ \alpha_i|{\bf h}_{i,j,k}^H{\bf w}'_i|^2\leq {\bf e}_{i,j,k}^T{\bf \Gamma}, \forall j\neq i, k ~~~~~~~~~~~~
\\\alpha_i>0~~~~~~~~~~~~~~~~~~~~~~~~~~~~~~~~~~~~~~~~~~~
\end{eqnarray}

The above scaling problem is a linear programming problem about the
one-dimension variable $\alpha_i$ and can be solved very fast.
Denote the minimum scaling ratio as $\alpha_i^\star$ and the
associated vector as ${\bf w}_i^\star$, then the corresponding beamformer is ${\bf w}_i=\sqrt{\alpha_i^\star}\cdot{\bf w}_i^\star$.
Both the randomization and scaling procedures are implemented
locally by each base station and do not break the distributed nature
of the algorithm. Notice although the proposed algorithm cannot guarantee the optimal solution
for the original problem ${\bf P}(\bm \gamma)$, it is seen from the simulation results in Section
V that this algorithm achieves the optimal solution of the original problem in most cases, especially
when the network size is small.

Finally, the decentralized algorithm is summarized below.

\vspace{0.4cm} \hrule \hrule \vspace{0.2cm} \textbf{Algorithm 1:
Decentralized algorithm for QoS beamforming} \vspace{0.2cm} \hrule
\vspace{0.3cm} ~~
\begin{itemize}
\item Step 1.~ Initialize the IT vector $\bf \Gamma$(0) with certain
values, e.g., ${\bf \Gamma}(0)=\bf 1$ and set the iteration number
$n=0$.
\item Step 2.~ Each base station locally solves its sub-problem \eqref{P_sub} and
broadcasts the resulting ${\bf g}_i$ (via the backhaul signaling) to
the other $N-1$ cells' base stations.
\item Step 3.~ Based on the reception of the subgradient vectors from all other cells, each base station updates the IT vector $\bf
\Gamma$(n) according to \eqref{Gamma_update}.
\item Step 4.~ set $n=n+1$ and go to Step 2 until meet the stopping
condition.
\item Step 5.~ For each base station $i$, if ${\bf W}_i^\star$ is rank-one,
then the optimal beamformer ${\bf w}_i$ is the eigenvector of ${\bf
W}_i^\star$; Otherwise the base station implements the randomization
and scaling procedures to get the near-optimal beamformer.
\end{itemize}
\vspace{0.2cm} \hrule \vspace{0.4cm}

The main information exchange in this decentralized beamforming
scheme is the real-valued subgradient vectors. During each
iteration, for each base station, it should broadcast its subgradient vector
${\bf g}_i$, which only has $N\times K$ nonzero entries. The sum signaling overhead among the
base stations in one iteration is thus $O\left(N^2K\right)$. Denote the number of iteration times as
$N_b$, then the total signaling is $O\left({N_bN^2K}\right)$. Through the simulations shown in Section V, we can see that the
algorithm converges very fast and can achieve major part of the
beamforming's gain after only several iterations. Furthermore, this
distributed algorithm can also work for the conventional
interference channel, which is a special case ($K=1$) of the
multi-cell multicast networks.

\section{Max-Min SINR Beamforming}
In this section, we study the beamforming problem for the MMS scheme
under individual base station power constraints.  The problem is
formulated as follows
\begin{eqnarray}
\label{S_p_ob}{\bf S}({\bf p}): &\max\limits_{\{{\bf w}_i\}_{i=1}^N}
\min\limits_{\forall i,\forall k}& \frac{|{\bf h}_{i,i,k}^H{\bf
w}_i|^2}{\sum\limits_{j\neq i}^N|{\bf h}_{j,i,k}^H{\bf
w}_j|^2+\sigma^2_{i,k}}\\& \mbox{s.t.}&\|{\bf w}_i\|^2\leq
p_i,\forall i
\end{eqnarray}
where ${\bf p}=[p_1, p_2,...,p_N]^T$ is the power constraint vector
for base-stations in the network.

\subsection{Connection with power optimization}
Define a weighted peak power minimization problem as
\begin{eqnarray}
\label{Q_gamma_p_ob}{\bf Q}(\gamma, {\bf p}):\min\limits_{\{{\bf
w}_i\}_{i=1}^N}\max\limits_{\forall i}~~\frac{1}{p_i}\|{\bf
w}_i\|^2~~~~~~~~~~~~~~~~~~
\\ \mbox{s.t.}~~\frac{|{\bf h}_{i,i,k}^H{\bf
w}_i|^2}{\sum\limits_{j\neq i}^N|{\bf h}_{j,i,k}^H{\bf
w}_j|^2+\sigma^2_{i,k}}\geq \gamma, \forall i,\forall k
\end{eqnarray}
where $\gamma$ is the common SINR target for all cells. It can be
seen that the two problems ${\bf S}({\bf p})$ and ${\bf Q}(\gamma,
{\bf p})$ are connected with each other through the power vector
$\bf p$. We use the notation $\gamma={\bf S}({\bf p})$ to stand for
the optimal objective value of problem ${\bf S}({\bf p})$, which
means that the maximum worst-case SINR for ${\bf S}({\bf p})$ is
$\gamma$. For the problem ${\bf Q}(\gamma, {\bf p})$, we denote the
associated optimum value as $p={\bf Q}(\gamma, {\bf p})$. Then the
following theorem tells the relationship between these two problem:

{\bf Theorem 2}: The SINR optimization problem of \eqref{S_p_ob} and
the power optimization problem of \eqref{Q_gamma_p_ob} are inverse
problems:
\begin{equation}\label{theorem2_1}
\gamma={\bf S}\left({\bf Q}(\gamma, {\bf p})\cdot{\bf p}\right)
\end{equation}
\begin{equation}\label{theorem2_2}
1={\bf Q}\left({\bf S}({\bf p}), {\bf p}\right).
\end{equation}

\begin{proof}
We first prove \eqref{theorem2_1} by contradiction. Let $p$ and
$\{{\bf w}_i\}_{i=1}^N$ be the optimal solution of ${\bf Q}(\gamma,
{\bf p})$, and $\widetilde{\gamma}$ and $\{{\bf
\widetilde{w}}_i\}_{i=1}^N$ be the optimal solution of ${\bf
S}(p\cdot{\bf p})$. We assume that $\widetilde{\gamma}\neq \gamma$.
Then if $\widetilde{\gamma}<\gamma$, then we can choose $\{{\bf
w}_i\}_{i=1}^N$ as the solution for ${\bf S}(p\cdot{\bf p})$, which
provides a larger objective value $\gamma$. This is a contradiction
of the optimality of $\{{\bf \widetilde{w}}_i\}_{i=1}^N$ for ${\bf
S}(p\cdot{\bf p})$. Otherwise, if $\widetilde{\gamma}>\gamma$, then
we can find a constant $c<1$ to scale the solution set $\{{\bf
\widetilde{w}}_i\}_{i=1}^N$ while still satisfying the SINR
constraints of problem ${\bf Q}(\gamma,\bf p)$. Since $\{{\bf
\widetilde{w}}_i\}_{i=1}^N$ satisfy the power constraints in ${\bf
S}(p\cdot{\bf p})$ which means that $\max_{\forall
i}\frac{1}{p_i}\|\widetilde{{\bf w}}_i\|^2=p$, the resulting set
$\{c{\bf \widetilde{w}}_i\}_{i=1}^N$ achieve a smaller objective
value (weighted peak base-station power) than $p$, which contradicts
the assumption that $\{{\bf w}_i\}_{i=1}^N$  is the optimal solution
of ${\bf Q}(\gamma, \bf p)$. Thus we must have $\widetilde{\gamma}=
\gamma$.

The proof of \eqref{theorem2_2} is similar and therefore omitted.
\end{proof}

In addition, we find numerically that the optimal objective values of both the
two problems are monotonically non-decreasing in the constraint
parameters $\bf p$ and $\gamma$. Such finding is reasonable since
with more power on each base station, the larger SINR can be
achieved, and vice versa.

A similar theorem on inverse property for single-cell multicasting has been proved in \cite{Karipidis}. Unlike \cite{Karipidis}, our max-min SINR problem for multicell multicasting cannot be solved by directly solving the corresponding QoS problem since each base station is subject to an individual power constraint. Instead, we show the this max-min-SINR problem with multiple power constraints can be solved by solving another weighted peak power minimization problem.

\subsection{Inversion-property-based algorithm}
Since the problem ${\bf S}(\bf p)$ is non-convex, similar to section
III, we apply the semidefinite relaxation and get the relaxed
problem ${\bf S}_1(\bf p)$ as below
\begin{eqnarray}
{\bf S}_1({\bf p}):  \max\limits_{\{{{\bf W}}_i\}_{i=1}^N}\min\limits_{\forall i,\forall
k}\frac{\mbox{Tr}\{{\bf H}_{i,i,k}{\bf W}_i\}}{\sum\limits_{j\neq
i}^N\mbox{Tr}\{{\bf H}_{j,i,k}{\bf W}_j\}+\sigma^2_{i,k}}\\
\mbox{s.t.}~~\mbox{Tr}\{{\bf W}_i\}\leq p_i,\forall i~~~~~~~~~~~~~~~~~~\\{\bf
W}_i\succeq{\bf 0},\forall i~~~~~~~~~~~~~~~~~~~~~~~~
\end{eqnarray}
The non-convex rank-one constraint has been dropped. We first get
the optimal solution of ${\bf S}_1(\bf p)$ based on Theorem 2. The
inverse problem of problem ${\bf S}_1(\bf p)$ is just the relaxed
version of ${\bf Q}(\gamma, \bf p)$, which is defined as follow
\begin{eqnarray}
{\bf Q}_1(\gamma, {\bf p}):\min\limits_{\{{\bf
W}_i\}_{i=1}^N}\max\limits_{\forall i} \frac{1}{p_i}\mbox{Tr}\{{\bf
W}_i\}~~~~~~~~~~~~~~~~~~~~~~~~~~
\\ \mbox{s.t.}~~{\mbox{Tr}\{{\bf H}_{i,i,k}{\bf
W}_i\}}\geq \gamma{\sum\limits_{j\neq i}^N\mbox{Tr}\{{\bf
H}_{j,i,k}{\bf W}_j\}+\gamma\sigma^2_{i,k}}, \forall i,\forall k
\\{\bf W}_i\succeq{\bf 0},\forall i~~~~~~~~~~~~~~~~~~~~~~~~~~~~~~~~~~~~~~~~~~~~~~~~
\end{eqnarray}
Introducing a slack variable $x$, we rewrite this problem in a more
elegant way
\begin{eqnarray}
{\bf Q}_1(\gamma, {\bf p}):\min\limits_{\{{\bf
W}_i\}_{i=1}^N,x}~~x~~~~~~~~~~~~~~~~~~~~~~~~~~~~~~~~~~~~~~~\\ \mbox{s.t.}~~{\mbox{Tr}\{{\bf H}_{i,i,k}{\bf
W}_i\}}\geq\gamma{\sum\limits_{j\neq i}^N\mbox{Tr}\{{\bf
H}_{j,i,k}{\bf W}_j\}+\gamma\sigma^2_{i,k}} , \forall i,\forall k\\
\frac{1}{p_i}\mbox{Tr}\{{\bf W}_i\}\leq x, \forall i~~~~~~~~~~~~~~~~~~~~~~~~~~~~~~~~~~~~~~~~\\{\bf
W}_i\succeq{\bf 0},\forall i~~~~~~~~~~~~~~~~~~~~~~~~~~~~~~~~~~~~~~~~~~~~~~~~
\end{eqnarray}

 Then we can solve ${\bf S}_1(\bf p)$ by iteratively
solving its inverse problem ${\bf Q}_1(\gamma, \bf p)$ for different
$\gamma$'s. Notice that ${\bf Q}_1(\gamma, \bf p)$ is a SDP problem
with strong duality and thus can be solved efficiently using the
interior method. Due to the inversion property, if $\gamma_0^\star$
is the optimal value for ${\bf S}_1(\bf p)$, then the optimal value
for the problem ${\bf Q}_1(\gamma_0^\star, \bf p)$ should be equal
to $1$. With the non-decreasing monotonicity, we can find the
optimal value $\gamma_0^ \star$ efficiently by a one-dimensional
bisection search over $\gamma$. When we get the optimal solution of
${\bf S}_1(\bf p)$, based on the resulting matrices $\{{\bf
W}_i^\star\}_i^N$, we apply the EVD or the randomization and scaling
to obtain the final beamformers for all the base stations. Finally,
the beamforming algorithm is summarized below

\vspace{0.4cm} \hrule \hrule \vspace{0.2cm} \textbf{Algorithm 2:
Inversion-property-based algorithm  for MMS beamforming}
\vspace{0.2cm} \hrule \vspace{0.3cm} ~~
\begin{itemize}
\item Step 1.~ Initialize the interval $[L,U]$ which contains the
optimal value $\gamma_0^\star$ of ${\bf S}_1( \bf p)$, e.g.,
$L=0,U=\max_{\forall i,\forall k}~p_i\cdot\|{\bf
h}_{i,i,k}\|^2/\sigma^2$, and set the iteration number $n = 0$.
\item Step 2.~ Set $\gamma=(L+U)/2$, and solve the problem ${\bf
Q}_1(\gamma, \bf p)$.
\item Step 3.~ If the optimal value $p_0^\star$ of ${\bf
Q}_1(\gamma, \bf p)$ is larger than $1$, set $U=\gamma$; Otherwise,
set $L=\gamma$.
\item Step 4.~ Set $n=n+1$ and go back to Step 2 until meet the
stopping condition.
\item Step 5.~ If ${\bf W}_i^\star$ is rank-1 for all $i$, we can obtain
the optimal solution for ${\bf S}(\bf p)$ by EVD; Otherwise, the
central controller does randomization and scaling to obtain the
approximate solution.
\end{itemize}
\vspace{0.2cm} \hrule \vspace{0.4cm}

Note that, in each iteration, we need to solve the
weighted peak power minimization problem, which is the main difference from
the algorithm in \cite{Karipidis}. Although this algorithm gives an approximate solution for ${\bf
S}(\bf p)$, from the simulation results in Section V, we can see
that it obtains the optimal solution in most cases.

\section{Simulation Results}
In this section, we provide numerical examples to illustrate the
performance of the proposed multicell multicast beamforming designs.
Within each cell, the channel is assumed as the normalized Rayleigh
fading channel, i.e., the elements of each channel vector are
independent and identically distributed circularly symmetric
zero-mean complex Gaussian random variables with unit variance. For
the intercell channels, i.e., the channels from the $i$th cell's
base station to the users in the $j$th cell ($j \neq i$), we
introduce the average large-scale fading ratio
$\varepsilon,0<\varepsilon<1$. A big $\varepsilon$ means large
intercell interference, which often occurs when the users are at the
boundary of each cell. Here, we set $\varepsilon=\frac{1}{2}$. A
common noise variance is set to be $\sigma^2_{i,k}=1,\forall i,k$.
Throughout this section, we use the notation ``$N-K-N_t$" to describe
the system configuration, which means that the network has $N$ cells
with $K$ mobile users in each cell and each base station has $N_t$
antennas. For all simulations, 200 channel realizations are
simulated and 100 Gaussian randomizations are generated if the
randomization method is needed.

\subsection{Performance comparison with existing schemes}
In this subsection, we consider the performance of the proposed
algorithms for the two transmission schemes. For the QoS scheme, we assume that the SINR targets for
the users in different cells are the same for simplicity. We first
illustrate the convergence of the algorithm. Fig. 2 plots the power
consumption during each iteration at different target SINR for the
$(2-2-4)$ network. The initial step size is $\mu=1$ and the
diminishing step size is $\mu(n)=\mu/\sqrt{(n)}$. Since the problem
${\bf P}_1(\bm \gamma)$ is convex, the proposed distributed
algorithm always converges to its optimal value. The simulation
result validates it. Also we can see that at the first few
iterations, the algorithm converges very fast and achieves the major
part of the optimal value.
\begin{figure}
\begin{centering}
\includegraphics[scale=0.45]{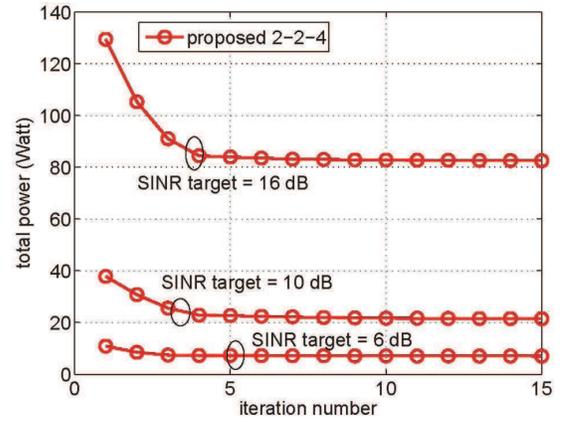}
\vspace{-0.1cm} \caption{Convergence behavior of the decentralized
algorithm for the QoS scheme.}
\label{fig:large_fading_multicell_iteration_process}
\end{centering}
\vspace{-0.3cm}
\end{figure}

\begin{figure}
\begin{centering}
\includegraphics[scale=0.45]{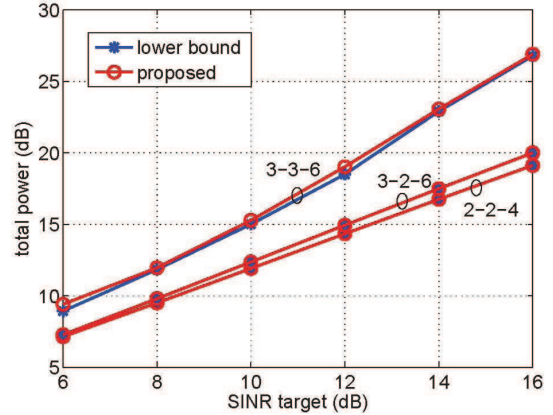}
\vspace{-0.1cm} \caption{Comparison of the proposed algorithm with
its lower bound for the QoS scheme.}
\label{fig:large_fading_multicell_iteration_process}
\end{centering}
\vspace{-0.3cm}
\end{figure}

\begin{figure}
\begin{centering}
\includegraphics[scale=0.45]{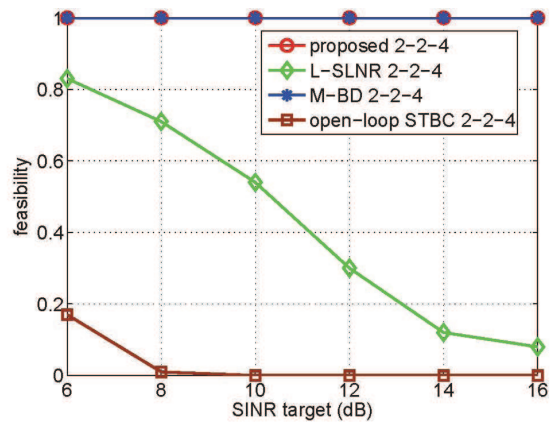}
\vspace{-0.1cm} \caption{Comparison on feasibility of the different
multicast beamforming algorithms for the QoS scheme.}
\end{centering}
\vspace{-0.3cm}
\end{figure}

In Fig. 3, we
illustrate the efficiency of the SDR approach by comparing it with
the lower bound, which is the solution of problem ${\bf
P}_1(\gamma)$ in \eqref{P_1_ob} where the rank-one constraint is
relaxed. It can be seen that although we take the semidefinite
relaxation, the proposed coordinated beamforming algorithm obtains
the optimal performance in most cases.

 We now compare the performance of the proposed SDR-based QoS beamforming in \textbf{Algorithm 1}
with two conventional multicell beamforming algorithms: multicell block
diagonalization (M-BD) \cite{Zhang1} and layered
signal-to-leakage-plus-noise ratio (L-SLNR) \cite{Zakhour}. Though
these two algorithms were proposed for multicell networks with
unicast traffic, they can be easily extended to multicasting. More specifically, for the M-BD algorithm, each base station chooses the beamforming vector which lies in the null space of the channels from the other $N-1$ base stations, i.e.,
\begin{equation}
{\bf w}_i\in\mbox{Null}\left(\hat{{\bf H}}_i\right)
\end{equation}
where $\hat{{\bf H}}_i=\left[{\bf h}_{1,i,1},...,{\bf h}_{1,i,K},...,{\bf h}_{i-1,i,1},...,{\bf h}_{i-1,i,K},\right.$ $\left.{\bf h}_{i+1,i,1},...,{\bf h}_{i+1,i,K},...,{\bf h}_{N,i,1},...,{\bf h}_{N,i,K}\right]^H$ and $\mbox{Null}(.)$ stands for the null space of a matrix. Then the QoS problem reduces to
a power allocation problem and we can solve it to obtain the resulting power for each base station. For the L-SLNR algorithm, each base station chooses
its beamforming vector as
\begin{eqnarray}\nonumber
{\bf w}_i\propto \mbox{max.} \mbox{eigenvector}\left(\left[\sum\limits_{j\neq i}^N\sum\limits_{l=1}^K{\bf h}_{i,j,l}{\bf h}_{i,j,l}^H+\sigma^2{\bf I}\right]^{-1}\right.\\\left.\cdot\left[\sum\limits_{l=1}^K{\bf h}_{i,i,l}{\bf h}_{i,i,l}^H\right]\right)~~~~~~~~~~~~~
\end{eqnarray}
which means that ${\bf w}_i$ should have the same direction as the eigenvector corresponding to the largest eigenvalue of the above matrix. We can similarly get the resulting beamforming vectors for the base stations by solving the reduced power allocation problem. As a performance benchmark, open-loop space time block coding (STBC) without requiring any CSI at each base station is also simulated.
Fig. 4 shows the comparison on feasibility \footnote{Although we
have assumed that the beamforming problem is feasible, each specific
algorithm may not be able to find the beamformers to meet the SINR
target due to its sub-optimality.}. Here, the feasibility percentage is obtained by
counting the number of trials among the 200 channel realizations
that each algorithm is able to find the solution to meet all the
constraints. It can be seen that both the open-loop STBC and the
L-SLNR beamformer almost do not work when SINR target $\gamma$
is large. This is expected as the open-loop STBC does not make any use of channel state information and serves purely as isotropic broadcasting. The L-SLNR beamformer, on the other hand, only tries to maximize the SLNR from the transmitter perspective and cannot guarantee the SINR maximization at the receiver end. From Fig. 4 it is also seen that the M-BD beamformer and the
proposed coordinated beamformer are always feasible in the considered
SINR target region. Here, the reason that the M-BD beamformer can work well is that the considered network is an interference-limited one and M-BD can null out all the interference for each user. The good performance of the proposed algorithm is expected as it can obtain the optimal solution in most cases under this particular setting (see Fig. 2). Therefore, in Fig. 5 we only compare
the energy efficiency of the proposed algorithm with M-BD algorithm. We can see
that the M-BD algorithm consumes much more power than the proposed
algorithm. In particular, at target SINR of $10$dB, M-BD consumes
$3$ dB more power in the $(2-2-4)$ system and $4$dB more power in
the $(3-2-6)$ system. Notice that the M-BD algorithm requires that
the number of transmitting antennas at each base station should be
larger than the total number of receive antennas at all users.
However, our proposed algorithm does not have this requirement.

\begin{figure}
\begin{centering}
\includegraphics[scale=0.45]{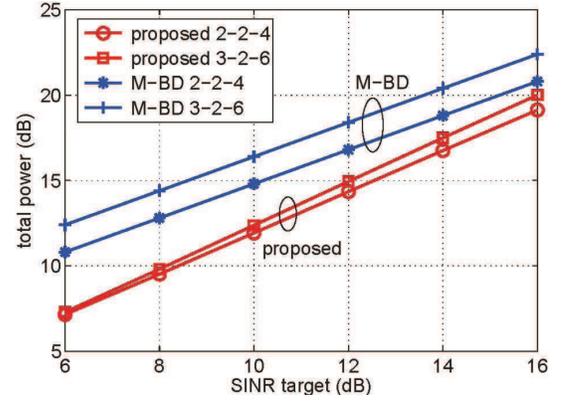}
\vspace{-0.1cm} \caption{Comparison on energy efficiency of the
different multicell multicast beamforming algorithms for the QoS
scheme.}
\end{centering}
\vspace{-0.3cm}
\end{figure}

For the max-min SINR beamforming scheme, we assume the power
constraint for each base station is the same. We first compare the
the performance of the proposed \textbf{Algorithm 2} with its upper bound in
Fig. 6. Similar to the QoS scheme, we can see that the gap from the
upper bound is very small and it achieves the optimal solution in
most cases. Fig. 7 compares its performance with the existing
algorithms. We can see that our proposed algorithm significantly
outperforms the other ones over a large range of individual power
constraint parameter. In particular, at the individual power
constraint of $10$dB in the $(3-2-5)$ system, the worst-case SINR
achieved by the proposed algorithm is $6$dB higher than L-SLNR,
 $8$dB higher than M-BD and $9$dB higher than open-loop STBC.
 From Fig. 7, it is also found that the M-BD algorithm performs the
 worst when the per base station power is small but is superior to the
 L-SLNR algorithm at large per base station power.

\begin{figure}
\begin{centering}
\includegraphics[scale=0.45]{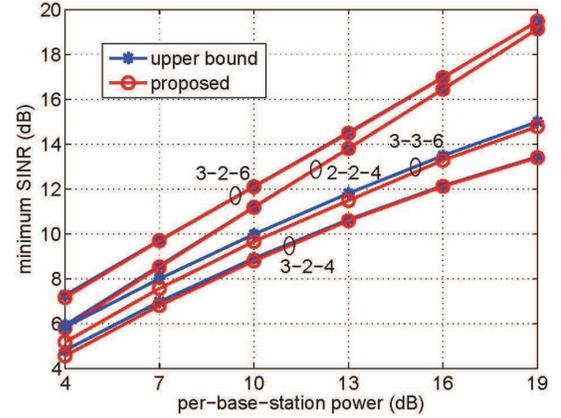}
\vspace{-0.1cm} \caption{Comparison on the minimum SINR of the
proposed algorithm with its upper bound for the MMS scheme.}
\label{fig:large_fading_multicell_iteration_process}
\end{centering}
\vspace{-0.3cm}
\end{figure}

\begin{figure}
\begin{centering}
\includegraphics[scale=0.45]{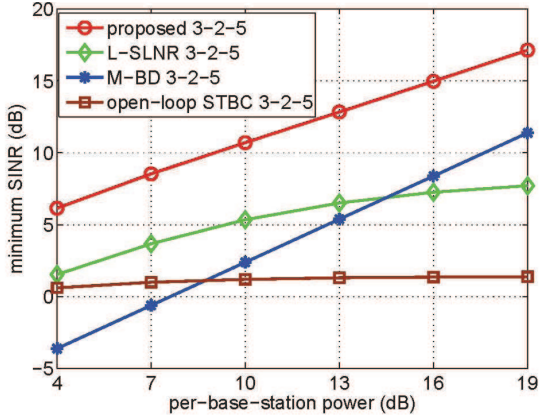}
\vspace{-0.1cm} \caption{Comparison of the different multicell
multicast beamforming algorithms for the MMS scheme.}
\label{fig:large_fading_multicell_iteration_process}
\end{centering}
\vspace{-0.3cm}
\end{figure}

\subsection{Effects of channel correlation}
In practical scenarios, the channels among different users may be
correlated with each other, especially for the users who are very
near to each other in the geographical position. In order to model
the correlated channels, we use the following Kronecker model
\cite{Mestre} \cite{Werner}.
\begin{equation}
{\bf H}_{kro}={\bf C}^{1/2}{\bf G},
\end{equation}
where ${\bf H}_{kro}$ is a channel matrix whose rows are correlated
with each other and ${\bf G}$ is a matrix with i.i.d. circularly
symmetric Gaussian entries with zero mean and unit variance (This is for
intracell channels by default.  If it is intercell channel, the variance is $\varepsilon^2$). We model channel correlation matrix $\bf C$ as a Hermitian
Toeplitz matrix with exponential entries $[{\bf C}]_
{i,j}=r^{|i-j|}$ \cite{Gore}. Here, $r$ can be seen as the correlation
ratio and $0\leq r \leq 1$.

We first investigate the effects of intercell-user channel
correlation on the feasibility of the QoS problem in Fig. 8. The
correlation ratio $r$ is set to be $0.5$, $0.7$ and $0.9$, where $r=0$
stands for the independent channel. We can see that when the
intercell-users' channels are correlated, the feasibility decreases,
which justifies our statement in Section III-A, i.e., the intercell-user channel correlation has a negative impact on the feasibility of the QoS problem.

Fig. 9 shows the effects of  intracell-user channel correlation for the QoS
beamforming scheme. The correlation ratio $r$ is also set to be 0.5, 0.7 and 0.9. From the
results, we can see that when the intracell-user channels are correlated,
it is helpful for the system. More specifically, the consumed power
becomes less. We have observed the similar results on the usefulness of intracell-user channel correlation
on the MMS scheme, which are ignored here due to page limit.
\begin{figure}
\begin{centering}
\includegraphics[scale=0.45]{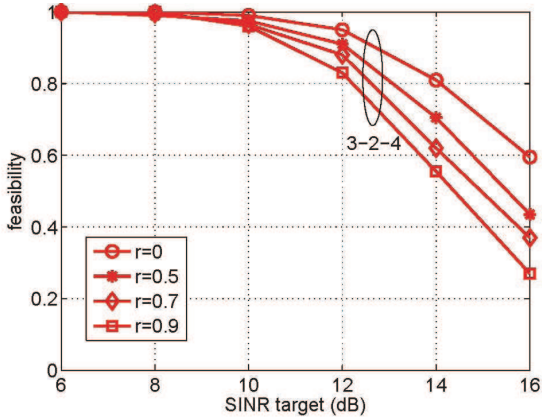}
\vspace{-0.1cm} \caption{Feasibility of the proposed algorithm with
correlated intercell-user channels for the QoS scheme.}
\label{fig:large_fading_multicell_iteration_process}
\end{centering}
\vspace{-0.3cm}
\end{figure}

\begin{figure}
\begin{centering}
\includegraphics[scale=0.45]{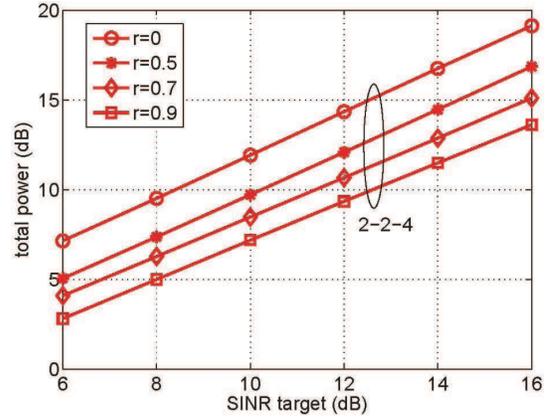}
\vspace{-0.1cm} \caption{Total transmitted power with correlated
intracell-user channels for the QoS scheme.}
\label{fig:large_fading_multicell_iteration_process}
\end{centering}
\vspace{-0.3cm}
\end{figure}

\section{Conclusion}
This paper considered two coordinated mulicast beamforming schemes for multicell networks. For the QoS scheme, we provided a necessary condition for the beamforming problem to be feasible when the
network shares a common SINR target and proposed a decentralized
algorithm to implement the coordinated beamforming.
For the max-min SINR scheme, we considered individual base station power constraints and
also proposed an efficient beamforming algorithm.
Besides, we also investigated the impacts of intercell-user and intracell-user
channel correlation on the multicast network.

There are several other issues to be further investigated for
multicell multicast beamforming. First, finding the sufficient
condition for the feasibility of the QoS problem remains open.
Second, a decentralized algorithm to implement the beamforming
design for the MMS scheme  is to be designed. Last but not least, it
is also interesting to consider robust beamforming when the channel
state information at each base station is not perfect.
\appendices
\section{Proof of Lemma 1}
\label{Proof P1} We first define the beamformer matrix ${\bf W}$
which is full-rank as below
\begin{equation}\label{W}
{\bf W}=\begin{bmatrix} {\bf w}_{1} & 0 & 0&  \cdots & 0 \\ 0 & {\bf
w}_{2} &0& \cdots & 0
\\\vdots & \cdots &~&~\\0 & 0 & 0&\cdots
& {\bf w}_{N} \end{bmatrix}~~~~~~~~~~~~~~~~~~~~~~~~~~~
\end{equation}
Based on the fact that $\mbox{SINR}_{i,k}\leq \mbox{SIR}_{i,k}$, we
have
\begin{eqnarray}\nonumber
{\mbox{SINR}_{i,k}}=\frac{{|{\bf h}_{i,i,k}^{\emph H}{\bf
w}_{i}|}^2}{\sum\limits^{N}_{j\neq i}{|{\bf h}_{j,i,k}^{\emph H}{\bf
w}_{j}|}^2+\sigma^2_{i,k}}~~~~~~\\\label{SINR_i_k}\leq\frac{{|{\bf h}_{i,i,k}^{\emph H}{\bf
w}_{i}|}^2}{\sum\limits^{N}_{j\neq i}{|{\bf h}_{j,i,k}^{\emph H}{\bf
w}_{j}|}^2}=\frac{1}{\frac{1}{\eta_{i,k}}-1}
\end{eqnarray}
where we define the $\eta_{i,k}$ as below
\begin{eqnarray}\nonumber
\eta_{i,k}=\frac{|{\bf h}_{i,i,k}^{\emph H}{\bf
w}_{i}|^2}{\sum\limits_{j=1}^{N}|{\bf h}_{j,i,k}^{\emph H}{\bf
w}_{j}|^2}~~~~~~~~~~~~~~~~~~~~~~~~~~~~~\\\label{eta}=\frac{|[{\bf H}_{k}{\bf
W}]_{i,i}|^2}{\sum\limits_{j=1}^{N}|[{\bf H}_{k}{\bf
W}]_{i,j}|^2}=\frac{|[{\bf H}_{k}{\bf W}]_{i,i}|^2}{|[{\bf
H}_{k}{\bf W}{\bf W}^{\emph H}{\bf H}^{\emph H}_{k}]_{i,i}|}
\end{eqnarray}
the second equation in \eqref{eta} is because \eqref{H_k} and
\eqref{W}. We also know the monotonicity of $f(x)=1/(1/x~-~1)$ when
$x<1$. Comparing it with \eqref{SINR_i_k}, we can bound the
$\mbox{SINR}_{i,k}$ through bounding the associate $\eta_{i,k}$. Let
the singular value decomposition(SVD) of ${\bf H}_{k}{\bf W}$ be
denoted as ${\bf H}_{k}{\bf W}={\bf U}_{k}
\boldsymbol{\Sigma}_{k}{\bf V}_{k}^{\emph H}$, where both ${\bf
U}_{k}$ and ${\bf V}_{k}$ are $K\times r_{k}$ quasi-unitary matrix,
i.e., their columns are orthogonal with each other.
$\boldsymbol\Sigma_{k}$ is an $r_{k} \times r_{k}$ diagonal matrix
whose elements are the singular values and $r_{k}=\mbox{rank}({\bf
H}_{k}{\bf W})$. Then
\begin{equation}
\eta_{i,k}=\frac{|{\bf u}_{k,i}^{\emph H}\boldsymbol\Sigma_{k}{\bf
v}_{k,i}|^2}{{\bf u}_{k,i}^{\emph H}\boldsymbol\Sigma_{k}^{2}{\bf
u}_{k,i}},~i=1~2\cdots N, k=1~2\cdots K
\end{equation}
where ${\bf u}_{k,i}$ and ${\bf v}_{k,i}$ are the $\emph i$th
columns of ${\bf U}_{k}$ and ${\bf V}_{k}$. According to
Cauchy-Schwarz inequality, we have
\begin{eqnarray}\nonumber
{|{\bf u}_{k,i}^{\emph H}\boldsymbol\Sigma_{k}{\bf v}_{k,i}|}^2\leq
{\|{\bf u}_{k,i}^{\emph H}\boldsymbol\Sigma_{k}\|}^2\cdot{\|{\bf
v}_{k,i}\|}^2~~~~~\\=\left({\bf u}_{k,i}^{\emph
H}\boldsymbol\Sigma_{k}^{2}{\bf u}_{k,i}\right)\left({\bf
v}_{k,i}^{\emph H}{\bf v}_{k,i}\right)
\end{eqnarray}
Since ${\bf v}_{k,i}^{\emph H}{\bf v}_{k,i}=[({\bf H}_{k}{\bf
W})^\dag({\bf H}_{k}{\bf W})]_{i,i}$, we conclude that
\begin{equation}
\eta_{i,k}\leq[({\bf H}_{k}{\bf W})^\dag({\bf H}_{k}{\bf W})]_{i,i}
\end{equation}
Thus we have
\begin{equation}
\begin{split}
\min_{i}\eta_{i,k}\leq
{\frac{1}{N}}\sum_{i=1}^{N}\eta_{i,k}\leq{\frac{1}{N}}\sum_{i=1}^{N}[({\bf
H}_{k}{\bf W})^\dag({\bf H}_{k}{\bf W})]_{i,i}\\
={\frac{1}{N}}\mbox{Tr}\{({\bf H}_{k}{\bf W})^\dag{\bf H}_{k}{\bf W}
\}=\frac{\mbox{rank}({\bf H}_{k}{\bf W})}{N}\leq
\frac{\mbox{rank}({\bf H}_{k})}{N}
\end{split}
\end{equation}
Further we can get
\begin{eqnarray}\nonumber
\min_{i,k}\eta_{i,k}=\min_{k}\left\{\min_{i}\eta_{i,k}\right\}~~~~~~~~~~~~~~~~~~~~\\\label{mini_eta}\leq
\min\{ \frac{\mbox{rank}({\bf H}_{1})}{N}, \frac{\mbox{rank}({\bf
H}_{2})}{N},\cdots,\frac{\mbox{rank}({\bf H}_{K})}{N}\}
\end{eqnarray}
So if the problem \eqref{P_gamma_ob} is feasible, then the minimum
SINR should be larger than the threshold $\gamma$. Plugging
\eqref{mini_eta} into \eqref{SINR_i_k}, we can conclude that
\begin{equation}
\gamma\leq \min_{i,k}{\mbox{SINR}}_{i,k}\leq
\left\{\frac{\mbox{rank}({\bf H}_1)}{{\emph N}-\mbox{rank}({\bf
H}_1)},\cdots,\frac{\mbox{rank}({\bf H}_K)}{{\emph
N}-\mbox{rank}({\bf H}_K)}\right\}
\end{equation}
which completes the proof of Lemma 1.

\section{Proof of Theorem 1}
We begin by computing the subgradient ${\bf g}_i$ from each
subproblem. The Lagrangian of the $i$th subproblem  is given by
\begin{eqnarray}\nonumber
{\emph L}({\bf W}_i,{\bf
\Gamma},{\boldsymbol{\lambda}})={\mbox{Tr}}\{({\bf
W}_i)\}~~~~~~~~~~~~~~~~~~~~~~~~~~~~~~~~~~\\\nonumber-\sum_{k=1}^{K}\lambda_{i,i,k}\left[\frac{1}{\gamma_i}{\mbox{Tr}}\{{\bf
H}_{i,i,k}{\bf W}_i\}-{\bf e}_{i,i,k}^{\emph H}{\bf
\Gamma}-\sigma^2_{i,k}\right]~~\\~+\sum_{j\neq
i}^{N}\sum_{k=1}^{K}{\lambda_{i,j,k}}\bigg[{\mbox{Tr}}\{{\bf
H}_{i,j,k}{\bf W}_i\}-{\bf e}_{i,j,k}^{\emph H}{\bf \Gamma}\bigg]~~~~~~~~~
\end{eqnarray}
Then the dual function $d_i({\bf \Gamma},{\boldsymbol{\lambda}})$ is
\begin{eqnarray}\nonumber
d_i({\bf \Gamma},{\boldsymbol{\lambda}})= \min\limits_{{\bf
W}_i}{\emph L}({\bf W}_i,{\bf
\Gamma},{\boldsymbol{\lambda}})~~~~~~~~~~~~~~~~~~~~~~~~~~~~~~~~\\=
\left(\sum\limits_{k=1}^{K}\lambda_{i,i,k}{\bf
e}_{i,i,k}^T-\sum\limits_{j\neq
i}^{N}\sum\limits_{k=1}^K\lambda_{i,j,k}{\bf e}_{i,j,k}^T\right){\bf
\Gamma}+f_i({\boldsymbol{\lambda}})
\end{eqnarray}
where
\begin{eqnarray}
\nonumber
f_i({\boldsymbol{\lambda}})~~~~~~~~~~~~~~~~~~~~~~~~~~~~~~~~~~~~~~~~~~~~
~~~~~~~~~~~~~~~~\\\nonumber =\min_{{\bf W}_i}\left({\mbox{Tr}}\{{\bf
W}_i\}-\sum_{k=1}^K\lambda_{i,i,k}\left[\frac{1}{\gamma_i}{\mbox{Tr}}\{{\bf
H}_{i,i,k}{\bf W}_i\}-\sigma^2_{i,k}\right] \right.\\ \left.
+\sum_{j\neq i}^N\sum_{k=1}^K\lambda_{i,j,k}{\mbox{Tr}\{{\bf
H}_{i,j,k}{\bf W}_i\}}\right)~~~~~~~~~~~~~~~~~~~~~~~~~
\end{eqnarray}
Since the problem ${\bf P}_i^{\mbox{sub}}(\gamma,{\bf \Gamma})$ is
convex, then the strong duality holds which means that
\begin{equation}
P_i^\star({\bf \Gamma})= \max_{{\boldsymbol{\lambda}}\succeq \bf
0}d_i({\bf \Gamma},{\boldsymbol{\lambda}})
\end{equation}
Denote ${\boldsymbol{\lambda}}^\star$ as the optimal Lagrange
multiplier for the dual problem, then we have
\begin{eqnarray}\nonumber
P_i^\star({\bf \Gamma})=d_i({\bf
\Gamma},{\boldsymbol{\lambda}}^\star)~~~~~~~~~~~~~~~~~~~~~~~~~~~~~~~~~~~~~~~~~~~~\\=\left(\sum_{k=1}^{K}\lambda_{i,i,k}^\star{\bf
e}_{i,i,k}^T-\sum_{j\neq i}^{N}\sum_{k=1}^K\lambda_{i,j,k}^\star{\bf
e}_{i,j,k}^T\right){\bf \Gamma}+f_i({\boldsymbol{\lambda}}^\star)
\end{eqnarray}
Define ${\bf g}_i$ as
\begin{eqnarray} {\bf
g}_i\triangleq\sum\limits_{k=1}^{K}\lambda_{i,i,k}^\star{\bf
e}_{i,i,k}-\sum\limits_{j\neq
i}^{N}\sum\limits_{k=1}^K\lambda_{i,j,k}^\star{\bf e}_{i,j,k},
\end{eqnarray}
we have
\begin{eqnarray}\nonumber
P_i^\star({\bf \Gamma})={\bf g}_i^H{\bf
\Gamma}+f_i({\boldsymbol{\lambda}}^\star)~~~~~~~~~~~~~~~~~~~~~~~~~~~~~~~~~~~~~~\\={\bf
g}_i^H({\bf \Gamma}-\widetilde{{\bf \Gamma}})+{\bf
g}_i^H\widetilde{{\bf \Gamma}}+f_i({\boldsymbol{\lambda}}^\star)\leq
{\bf g}_i^H({\bf \Gamma}-\widetilde{{\bf
\Gamma}})+P_i^\star(\widetilde{{\bf \Gamma}})
\end{eqnarray}
which is equivalent to that
\begin{equation}
P_i^\star(\widetilde{{\bf \Gamma}})\geq P_i^\star({\bf \Gamma})+{\bf
g}_i^H(\widetilde{{\bf \Gamma}}-{\bf \Gamma}).
\end{equation}
It means that ${\bf g}_i$ is the subgradient of $P_i^\star({\bm
\Gamma})$ and obtained for the $i$th subproblem.

In the same way, we can compute the global subgradient $\bf g$ of $
{\bf P}^{\mbox{mas}}(\bm\gamma,{\bf \Gamma})$ as below
\begin{eqnarray}\nonumber
{\bf g} &
=&\sum\limits_{i=1}^{N}\sum\limits_{k=1}^{K}\lambda_{i,i,k}^\star{\bf
e}_{i,i,k}-\sum\limits_{i=1}^{N}\sum\limits_{j\neq
i}^{N}\sum\limits_{k=1}^{K}\lambda_{i,j,k}^\star{\bf e}_{i,j,k} \\\nonumber ~
&
=&\sum\limits_{i=1}^{N}\left(\sum\limits_{k=1}^{K}\lambda_{i,i,k}^\star{\bf
e}_{i,i,k}-\sum\limits_{j\neq
i}^{N}\sum\limits_{k=1}^{K}\lambda_{i,j,k}^\star{\bf
e}_{i,j,k}\right) \\ ~ & =&\sum\limits_{i=1}^{N}{\bf g}_i
\end{eqnarray}
which completes the proof of Theorem 1.


\begin{thebibliography}{99}
\bibitem{Sidiropoulos} N. D. Sidiropoulos, T. N. Davidson, and Z.-Q.
Luo, ``Transmit beamforming for physical-layer multicasting," {\sl
IEEE Transactions on Signal Processing}, vol. 54, no. 6,  Jun 2006.

\bibitem{Karipidis} E. Karipidis, N. D. Sidiropoulos, and Z.-Q. Luo,
``Quality of service and max-min fair transmit beamforming to
multiple cochannel multicast groups," {\sl IEEE Transactions on
Signal Processing}, vol. 56, no. 3, Mar, 2008.


\bibitem{Ntranos} V. Ntranos, N. D. Sidiropoulos and L. Tassiulas,
``On multicast beamforming for minimum outage," {\sl IEEE
Transactions on Wireless Communications}, vol. 8, no. 6, Jun 2009.

\bibitem{Park} S. Y. Park and D. J. Love, ``Capacity limits of multiple
antenna multicasting using antenna subset selection,"
{\sl IEEE Transactions on Signal Processing}, vol. 56, no. 6, Jun
2008.

\bibitem{Motorola} Motorola Inc., ``Long term evolution (LTE): A
technical overview," Technical White Paper,http://business.motorola.com
/experienceltr/pdf/LTE\%20Technical\%20overview.pdf

\bibitem{Lozano} A. Lozano, ``Long-term transmit beamforming for
wireless multicasting," in {\sl Proc. ICASSP' 07}, April 2007,
Honolulu, Hawaii.

\bibitem{Shamai} S. Shamai(Shitz) and B. M. Zaidel, ``Enhancing the
cellular downlink capacity via co-processing at the transmitting
end," in {\sl Proc. IEEE Vehicle Technology Conference.(VTC)}, May
2001, vol. 3, pp. 1745-1749.

\bibitem{Dai} H. Zhang and H. Dai, ``Cochannel interference
mitigation and cooperative processing in downlink multicell
multiuser MIMO networks," {\sl EURASIP J. Wireless Commun. Netw.},
no. 2, 2004.

\bibitem{Jorswieck} R. Mochaourab and E. Jorswieck, ``Optimal beamforming
in interference networks with perfect local channel information," {\sl IEEE Transactions on
Signal Processing}, vol. 59, no. 3, March 2011.

\bibitem{Palomar} D. P. Palomar, M. Chiang, ``A tutorial on
decomposition methods for network utility maximization," {\sl IEEE
Jouranl on Selected Areas in Communications}, vol. 21, no. 8, Aug
2006.


\bibitem{Jordan} M. Jordan, X. Gong, G. Ascheid, ``Multicell Multicast
Beamforming with Delayed SNR Feedback," {\sl Proc. GLOBECOM 09},
Nov, 2009.

\bibitem{Dartmann} G. Dartmann, X. Gong, G. Ascheid, ``Low Complexity
Cooperative Multicast Beamforming in Multiuser Multicell Downlink
Networks", {\sl Proc. CROWNCOM' 2011}, Jun, 2011.

\bibitem{Dartmann1} G. Dartmann, X. Gong, G. Ascheid, ``Cooperative Beamforming
with Multiple Base Station Assignment Based on Correlation
Knowledge", {\sl Proc. VTC}, 2010.

\bibitem{Wiesel} A. Wiesel, Y. C. Eldar, S. Shamai (Shitz), ``Linear
precoding via conic optimization for fixed MIMO receivers," {\sl
IEEE Transactions on Signal Processing}, vol. 54, no. 1, Jan 2006.

\bibitem{Boyd1} M. Grant and S. Boyd. {\sl CVX: Matlab software for disciplined convex
programming} [online]. Available: http://stanford.edu.boyd/cvx.

\bibitem{S.Zhang} S. Zhang, Y. Huang, ``Complex quadratic
optimization and semidefinite programming," {\sl SIAM J. Optim.},
vol. 16, no. 3, Jan, 2006.

\bibitem{Dahrouj} H. Dahrouj, W. Yu, ``Coordinated beamforming for
the multicell multi-antenna wireless system," {\sl IEEE Transactions
on Wireless Communications}, vol. 9, no. 5, May 2010.

\bibitem{Zhang} R. Zhang, S. Cui, ``Cooperative interference
management with MISO beamforming," {\sl IEEE Transactions on Signal
Processing}, vol. 58, no. 10, Oct 2010.


\bibitem{Zhang1} R. Zhang, ``Cooperative multi-cell block diagonalization with
per-base-station power constraints," {\sl IEEE Jouranl on Selected
Areas in Communications}, vol. 28, no. 9, Dec. 2010.

\bibitem{Zakhour} R. Zakhour and D. Gesbert ``Distributed multicell-MISO precoding
 using the layered virtual SINR framework," {\sl IEEE Transactions
on Wireless Communications}, vol. 9, no. 8, Aug. 2010.


\bibitem{Mestre} X. Mestre and J. Fonollosa. ``Capacity of MIMO channels:
Asymptotic evaluation under correlated fading," {\sl IEEE Jouranl on
Selected Areas in Communications}, vol. 21, no. 5, June 2003.

\bibitem{Werner} K. Werner, M. Janson and P. Stoica. ``On estimation of covariance matrices with kronecher product
structure," {\sl IEEE Transactions on Signal Processing}, vol. 56,
no. 2, Feb 2008.

\bibitem{Gore} A. Paulraj, R. Nabar, and D. Gore, {\sl Introduction to Space-Time
Wireless Communications}. Cambridge, U.K.: Cambridge Univ. Press,
2003
\end{thebibliography}
\end{document}